\documentclass[aps,prb,twocolumn]{revtex4-2}
\usepackage{graphicx}
\usepackage{epsfig}
\usepackage{amsmath}
\usepackage{graphics}
\usepackage{epsfig}
\usepackage{bm}
\usepackage[utf8x]{inputenc}
\usepackage{tikz}
\usetikzlibrary{calc}

\begin{document}

\title{Molecular electronic junctions with  stochastic structural transitions}
\author{Daniel S. Kosov}
\address{College of Science and Engineering,  James Cook University, Townsville, QLD, 4811, Australia }


\begin{abstract}
We present a theoretical approach to include dynamical aspects of molecular  structural fluctuations, such as, for example, forming and breaking hydrogen bonds, isomerizations, and dynamical supramolecular structures, in nonequilibrium Green's functions electron transport calculations.
Structural transitions are treated as a stochastic telegraph process, and
the primary quantity of interest is a retarded Green's function averaged over realizations of a stochastic process. Using the Novikov-Furutsu functional stochastic calculus method, we derive equations of motion for stochastically averaged retarded Green's function
in closed form. Consequently, we obtain the expression for electric current averaged over transitions, which depends not only on probabilities of observing particular molecular structures but also on the dynamics. However, the proposed method has a significant limitation - we have to assume that the imaginary parts of retarded self-energies produced by left and right electrodes are proportional to each other; this significantly restricts possible applications of the  theory. Several examples illustrate the proposed approach.
\end{abstract}

\maketitle

\section{Introduction}
Fluctuations of geometry are often present in molecular electronic junctions. 
Bistable molecular conformations\cite{doi:10.1021/nl034710p,Baber2008,Donhauser2001,Auwarter2011,Cho2018a,noise14}, forming and breaking molecule-electrode chemical bonds\cite{Nichols2010} have been observed experimentally. The stochastic switching between several molecular junction structures is ubiquitous in the emerging field of (supra) molecular electronics, where host-guest interactions, hydrogen bonding, $\pi-\pi$ interactions, and non-covalent interactions connect mechanically interlocked molecules\cite{Chen:2021ud}.

If stochastic structural changes are slow compared to electron tunneling time, the problem becomes trivial. One simply  must perform standard static electron transport calculations for several frozen molecular geometries and average the result. However, if the stochastic switching rate becomes comparable with the rate for electron transfer, the problem becomes dynamic, and the proper treatment is highly nontrivial. The development of a practical theory for electron transport, which considers the dynamics of stochastic switching, is our goal here.

We assume that the molecular junction switches randomly between two static molecular Hamiltonians with some constant switching rate as time progresses. That means we take that the telegraph process describes stochastic dynamics.
The theoretical literature on quantum transport with telegraphic switching Hamiltonians exists but the scope is  limited. In the 1990s, Galperin et al. studied average transparency\cite{telegraph94} and low-frequency noise\cite{telegraph95} through double barriers with dynamic impurity. Entin-Wohlman et al.\cite{telegraph17}used nonequlibrium Green's functions (NEGF) to study quantum heat transport via a fluctuating electronic level, proposed as a model for an applied stochastic electric field. Gurvitz et al. \cite{gurvitz16} also used a fluctuating electronic level and  analyzed steady-state along with transient dynamics. We have recently investigated telegraph noise in a junction with electron-phonon and electron-electron interactions within the master equation framework by adding a stochastic component into the Liouvillian\cite{kosov18} or  introducing additional rates\cite{PhysRevB.100.235430} into quantum master equations.

The master equation method\cite{kosov18,PhysRevB.100.235430} to deal with the problem suffered from the common deficiencies of the master equation approach to quantum transport, such as the Markovian assumption, weak electrode-molecule coupling, and the lack of level broadening. Previous NEGF approaches\cite{telegraph94,telegraph95,telegraph17} treated the problem within the wide-band approximation and restricted the treatment to a single resonant level. These approximations severely limit the applications but simplify the treatment considerably. For example, the imaginary part of retarded self-energy becomes a delta-function within the wide-band approximation enabling the simple mathematical handling of stochastic equations of motion for Green's function, and the use of a single level trivializes the calculation of time-ordered exponents. 

Here, we develop NEGF-based electron transport theory without resorting to the wide-band treatment of electrodes nor making any assumption about the size or structure of the molecular Hamiltonian. 
We present the expression for electric current averaged over stochastic process associated with molecular transitions in a form ready to be used for practical calculations of large molecular electronic junctions. However, we must admit that our approach is incomplete; it suffers from a limitation (though less severe than in the methods previously developed) - we have to assume that level broadening functions of the left and right electrodes are proportional to each other. This limitation stems from our inability to compute averaged over the stochastic process lesser Green's function - the reason for this will be made clear later in the paper.

We use atomic units for quantum transport  throughout the paper.
\section{Theory}
\subsection{Hamiltonian with telegraphic switching between two molecular isomers}
The  Hamiltonian of the molecular junction consists of  Hamiltonians for molecule $H_M$, left  and right electrodes, $H_L$ and $H_R$, and the interaction which couples the molecule to left and right electrodes, $H_{ML}$
and $H_{MR}$:
\begin{equation}
H(t)=H_M(t) +  H_L + H_R + H_{ML} + H_{MR}.
\label{hamiltonian}
\end{equation}
Suppose that  the molecule can switch between two states A and B described by different Hamiltonians
$H_A$ and $H_B$. 
These  transitions between A and B  make the molecular Hamiltonian explicitly time-dependent $H_M(t)$.
We write Hamiltonians for the states A and B  in the following general  forms:
\begin{equation}
H_{A/B}(t) = \sum_{ij}  H^{A/B}_{ij}  a^\dag_i a_j,
\label{ha}
\end{equation}
where
$H^{A/B}_{ij}$  are matrix elements computed in molecular single-particle states and $a^{\dagger}_i$($a_i $)  creates (annihilates) an electron  in the molecular single-particle states  $i$ in the molecule.
To cast these Hamiltonians in the form of stochastic telegraphic process we write matrices $\bm H^{A/B}$ as
\begin{equation}
\bm H^A = \bm H + \bm V
\end{equation}
and 
\begin{equation}
\bm H^B = \bm H - \bm V.
\end{equation}
This enables us to write the time-dependent molecular Hamiltonian in the following form
\begin{equation}
H_M(t) = \sum_{ij} [ H_{ij} + \xi(t)  V_{ij} ] a^\dag_i a_j,
\label{hm}
\end{equation}
where  function $ \xi(t)$  describes  stochastic telegraph  progress;  it switches stochastically between  two states as the time progresses
\begin{equation}
\xi(t)=  x (-1)^{n(0,t)},
\end{equation}
where $n(t,t')$ is the random sequence of integer numbers describing the number of switches on the time interval $[t,t']$. 
We assume that the initial isomer is not known, the molecule  have equal probabilities to be observed in state A or B.  Therefore, 
$x$ is random variable with distribution
\begin{equation}
\label{pa}
p(x) = \frac{1}{2} (\delta(x-1) + \delta(x+1))
\end{equation}

We assume that the distribution of times at which the transitions $A \leftrightarrow B$ occur is the Poisson stationary distribution,  hence
the probability to observe $n$ transitions during the time interval $[t,t']$ is
\begin{equation}
p_{n(t,t')=n} = \frac{{\overline{n(t,t')}~}^n}{n!} e^{-\overline{n(t,t')}}
\label{pn}
\end{equation}
with average number of telegraph jumps during the time interval
\begin{equation}
\overline{n(t,t')} =  \sum_{n=0}^{\infty} n p_{n(t,t')=n}  = k_{AB}  (t-t'),
\end{equation}
where $k_{AB}$ is  the average number of transitions per unit time - transition rate.
Using the distribution (\ref{pn}), we compute the first few moments of the stochastic function $\xi(t)$ averaging over the realisations of the stochastic process.
\begin{equation}
\Big\langle \xi(t) \Big\rangle_\xi  = 0
\end{equation}
where $\Big\langle ... \Big\rangle_\xi $ means the averaging over all possible realisation of stochastic process $\xi(t)$ as well as the aditional averaging over distribution $p(x)$.
Two-time  correlation function computed for  $t_1 \ge t_2$ is 
\begin{multline}
\Big\langle \xi(t_1)  \xi(t_2)\Big\rangle_\xi  =  \Big\langle x (-1)^{n(0,t_1)}   x (-1)^{n(0,t_2)} \Big\rangle_\xi \\
=   \Big\langle x^2 (-1)^{n(t_2,t_1)}   \Big\rangle_\xi =  e^{-2 k_{AB} (t_1-t_2)},
\end{multline}
and all higher order correlation functions can be factorised if they are preliminary time-ordered ($t_1\ge t_2 \ge  t_3 \ge ... \ge t_N$)
\begin{multline}
\Big\langle \xi(t_1)  \xi(t_2) \xi(t_3) ... \xi(t_N) \Big\rangle_\xi  \\
= \Big\langle \xi(t_1)  \xi(t_2) \Big\rangle_\xi \Big\langle  \xi(t_3) ... \xi(t_N) \Big\rangle_\xi.
\end{multline}
We refer to the textbook on stochastic dynamical systems which contains many details relevant to our discussion here\cite{klyatskin2011}.

The electrodes are represented by non-interacting Hamiltonians with continuum spectrum
\begin{equation}
H_L+ H_R=  \sum_{ k\alpha}  \varepsilon_{k\alpha} a^{\dagger}_{k\alpha} a_{k\alpha},
\label{hl}
\end{equation}
the coupling between molecule and electrode is giving by tunneling interaction
\begin{equation}
H_{ML} + H_{MR}=
\sum_{k \alpha i } (t_{k\alpha i}  a^{\dagger}_{k \alpha} a_i +\mbox{h.c.} ).
\label{hlm}
\end{equation}
Here $a^{\dagger}_{ k\alpha}$($a_{ k\alpha}$) creates (annihilates) an electron in the single-particle state with energy $ \varepsilon_{k\alpha}$ of either the left ($\alpha=L$) or  the right ($\alpha=R$) electrodes, and  $t_{k\alpha i} $ tunneling amplitudes between single-particle states in electrodes and molecule.

\subsection{Green's functions and self-energies}
The exact (computed with full time-dependent stochastic Hamiltonian) molecular space Green's functions are
\begin{equation}
G _{ij}^A(t,t') = i \theta(t'-t) \langle \{a_i (t), a^\dag_j (t')\} \rangle,
\end{equation}
\begin{equation}
 G_{ij}^R (t,t') = -i \theta(t-t') \langle \{a_i (t), a^\dag_j (t')\} \rangle,
 \label{gr}
\end{equation}
\begin{equation}
 G_{ij}^<(t,t') = i \langle a^\dag_j (t')a_i (t) \rangle.
 \label{gl}
\end{equation}

The self-energies of electrodes are not affected by the time-dependence of the  molecular Hamiltonian, therefore, they are defined in standard way\cite{haug-jauho}.  Left and right retarded self-energies are
\begin{equation}
{\Sigma}_{\alpha ij}^R(t,t') = -i \theta(t-t') \sum_{k} t^*_{k\alpha i}  e^{-i \epsilon_{k\alpha} (t-t')} t_{l  j}.
\end{equation}
The advanced  and retarded self-energies are related to each other  via Hermitian  conjugation:
\begin{equation}
{\Sigma}_{\alpha ij }^A(t,t') = \Big( {\Sigma}_{\alpha ji}^R(t',t) \Big)^*.
\label{sigmaAt}
\end{equation}
The lesser self-energies are
\begin{equation}
{\Sigma}_{\alpha ij}^<(t,t') =2 \pi i \sum_{\alpha} t^*_{k \alpha i}  f_L(\epsilon_{k\alpha}) e^{-i \epsilon_{k\alpha} (t-t')} t_{l  j}
\end{equation} 
where  $f_\alpha$ are the Fermi-Dirac occupation numbers for electrode $\alpha$.
The total self-energies are the sum of contributions from the left and right electrodes 
\begin{equation}
{\Sigma}^{R,A,<}_{ij}(t,t') =  {\Sigma}^{R,A,<}_{L ij}(t,t')+ {\Sigma}_{R ij}^{R,A,<}(t,t').
\end{equation}

The retarded self-energies in the energy domain  are defined in the usual way as a Fourier transformation of time domain self-energies defined above:
\begin{equation}
\Sigma_{\alpha ij}^R(\omega)= \Delta_{\alpha ij }(\omega)  -\frac{i}{2} \Gamma_{\alpha ij}(\omega),
\end{equation}
where the level-width functions are
\begin{equation}
 \Gamma_{\alpha ij } (\omega) = 2 \pi \sum_{k} \delta(\omega-\epsilon_{k\alpha})  t^*_{k\alpha i} t_{k \alpha j}.
\end{equation}
The level-shift functions $\Delta_{L/R ij }(\omega)$ can be computed from $\Gamma_{L/R ij}(\omega)$  via Kramers-Kronig relation\cite{haug-jauho}. 
The advanced  self-energy is computed from the retarded self-energy as
\begin{equation}
\Sigma_{\alpha ij}^A(\omega) = (\Sigma_{\alpha ji}^R(\omega))^*,
\end{equation}
 and, then the lesser self-energy calculated using advanced and retarded self-energies
\begin{equation}
 \Sigma_{\alpha ij}^<(\omega)= f_{\alpha}(\omega)\left( \Sigma_{\alpha ij}^A(\omega) - \Sigma_{\alpha ij}^R(\omega)\right)= i f_{\alpha}(\omega) \Gamma_{\alpha ij}(\omega).
\end{equation}

\subsection{Averaging of retarded Green's function over the stochastic  process}
Our calculations of electric current in section (II-D)  will require the knowledge of retarded Green's functions only.
Therefore,
we begin with equation of motion for the retarded Green' s function:
\begin{multline}
( i \partial_t -   \bm  H  - \xi(t) \bm V  )  \bm  G^{R}(t,t') 
\\
-\int  dt_1   \bm   \Sigma^{R}(t,t_1)     \bm  G^{R}(t_1,t') = \delta(t-t'),
\label{em1}
\end{multline}
the matrix multiplication is implied here and below.
Next, we average this equation of motion over  the stochastic  process. 
The averaging  gives
\begin{multline}
( i \partial_t -   \bm  H  )  \langle  \bm    G^R(t,t') \rangle_\xi - \bm V \langle  {\xi}(t)   \bm    G^R(t,t') \rangle_\xi 
\\
-\int  dt_1   \bm   \Sigma^R(t,t_1) \langle   \bm    G^R(t_1,t') \rangle_\xi 
= \delta(t-t').
\label{de2}
\end{multline}
In computing the average we took into account that self-energy does not depend on stochastic process $\xi(t)$.
Alas this equation ({\ref{de2}) is not in the closed form to be solved since the average $ \langle \xi(t)\bm    G^R(t,t') \rangle_\xi$ is not known.
Let us evaluate it.
We notice that  the  Green's function can be considered as a functional of the stochastic process $\xi(t)$ 
\begin{equation}
  \bm    G^R(t,t')=   \bm    G^R[\xi](t,t').
\end{equation}
Using Shapiro-Loginov differential formula in stochastic calculus\cite{shapiro-loginov} (derivations for the derivative of $G^R$ are shown in  appendix B) gives
\begin{equation}
(\partial_t + 2 k_{AB}) \langle \xi(t)    \bm    G^R(t,t') \rangle_\xi =    \langle \xi(t)  \;\partial_t   \bm    G^R(t,t') \rangle_\xi.
\label{eqa}
\end{equation}
Substitution $\partial_t   \bm    G^{R}(t,t')$ from the initial equation of motion (\ref{em1}) into rhs of (\ref{eqa})
gives
\begin{multline}
(i \partial_t + 2 i k_{AB} -\bm   H   ) \langle \xi(t)    \bm    G^{R}(t,t') \rangle_\xi 
\\
- \int  dt_1   \bm   \Sigma^{R}(t,t_1)  \langle \xi(t)     \bm    G^{R}(t_1,t') \rangle_\xi=   \bm   V  \langle      \bm    G^{R}(t,t') \rangle_\xi
\label{eqB}
\end{multline}
Since $\Sigma^{R}(t,t_1)$ decays very quickly as $t$ deviates from $t_1$, for example, in the extreme case of wide band approximation $\Sigma^{R}(t,t_1) \sim \delta(t-t_1)$, it is safe to approximate the integral as
\begin{multline}
\int  dt_1   \bm   \Sigma^{R}(t,t_1)  \langle \xi(t)     \bm    G^{R}(t_1,t') \rangle_\xi \\
\simeq \int  dt_1   \bm   \Sigma^{R}(t,t_1)  \langle \xi(t_1)     \bm    G^{R}(t_1,t') \rangle_\xi 
\end{multline}
The result is  the system of two integro-differential equations
\begin{equation}
\left.
\begin{array}{l}
( i \partial_t -   \bm  H    )  \bm   {\mathcal G}^{R}(t,t') 
-  \bm V \bm {\mathcal F}^{R}(t,t') 
\\
-\int  dt_1   \bm   \Sigma^{R}(t,t_1)     \bm  {\mathcal G}^{R}(t_1,t') = \delta(t-t')
\\
\\
(i \partial_t + 2 i k_{AB} -\bm   H  )  \bm   {\mathcal F}^{R}(t,t') 
\\
- \int  dt_1   \bm   \Sigma^{R}(t,t_1)  \bm {\mathcal F}^{R}(t,t') =   \bm   V        \bm {\mathcal G}^{R}(t,t')
\end{array}
\right\}.
\label{2eq}
\end{equation}
Here we introduced the following notations for stochastically averaged quantities
\begin{equation}
\bm {\mathcal G}^R(t,t') =  \langle    \bm    G^R(t,t') \rangle_\xi,
\end{equation}
\begin{equation}
\bm {\mathcal F}^R(t,t') =  \langle \xi(t)     \bm    G^R(t,t') \rangle_\xi.
\end{equation}

The stochastically averaged retarded Green's function $\bm {\mathcal G}^R(t,t')$ depends on relative time only,  and  we see from the first equation in the system of equations (\ref{2eq}) that 
$\bm {\mathcal F}^R(t,t')$ must depend on relative time only too. Therefore, we can conveniently employ the Fourier transform to remove integrals and time derivatives, and  bring  (\ref{2eq}) to the form
\begin{equation}
\left.
\begin{array}{c}
( \omega -   \bm  H   -\bm   \Sigma^{R}(\omega)   )  \bm   {\mathcal G}^{R}(\omega) -  \bm V \bm {\mathcal F}^{R}(\omega) 
 = \bm I
\\
\\
(\omega -\bm   H  -   \bm   \Sigma^{R}(\omega) + 2 i k_{AB} )  \bm   {\mathcal F}^{R}(\omega) 
=   \bm   V        \bm {\mathcal G}^{R}(\omega)
\end{array}
\right\}
\end{equation}
Matrix $\bm {\mathcal F^R}(\omega)$ can be  eliminated from the system of equations resulting into the following expression for the retarded Green's function
\begin{multline}
\bm   {\mathcal G}^{R}(\omega)
\\
=\Big[  \omega -   \bm  H   -\bm   \Sigma^{R}(\omega)   - \bm V \frac{1}{\omega  -\bm   H  -   \bm   \Sigma^{R}(\omega)+ 2 i k_{AB} }   \bm V \Big]^{-1}
\label{gr}
\end{multline}

Let's check the  limiting cases. If $k_{AB} \rightarrow \infty$, the junctions switches infinitely fast between conformations and a tunneling electron does not see the presence of the stochastic potential $\xi(t) \bm V$ at all. Indeed, in this case 
\begin{equation}
\bm V \frac{1}{\omega -\bm   H  -   \bm   \Sigma^{R}(\omega) + 2 i k_{AB}  }   \bm V  \rightarrow 0
\end{equation}
in (\ref{gr}) and, as we have expected,
the retarded Green's function is reduced to the static expression for the junction described by time-independent Hamiltonian $\bm H$
\begin{equation}
\bm   {\mathcal G}^{R}(\omega)=\Big[  \omega -   \bm  H   -\bm   \Sigma^{R}(\omega)   \Big]^{-1}.
\end{equation}
In the opposite limit $k_{AB}=0$, the junction does not undego any telegraphic switching between confirmations, however, the averaging is still performed over initial state using distribution (\ref{pa}). Using stanard matrix algebra we get
\begin{multline}
\bm   {\mathcal G}^{R}(\omega)=\Big[  \omega -   \bm  H   -\bm   \Sigma^{R}(\omega)   - \bm V \frac{1}{\omega  -\bm   H  -   \bm   \Sigma^{R}(\omega) }   \bm V \Big]^{-1}
\\
=\frac{1}{2}\Big(\frac{1}{  \omega -   \bm  H -\bm V   -\bm   \Sigma^{R}(\omega)} +\frac{1}{ \omega -   \bm  H +\bm V   -\bm   \Sigma^{R}(\omega) }\Big),
\end{multline}
which again is consistent with our physical expectations.

\subsection{Current}

The general, exact  expression for electric current from $\alpha$ electrode into the molecule is\cite{haug-jauho}
\begin{multline}
J_\alpha(t)= 
2  \;  \text{Re} \int dt' \text{Tr} \Big[  \bm G^<(t, t') \bm \Sigma^A_{\alpha}(t',t)\\
+  \bm G^R(t, t')  \bm \Sigma^<_{\alpha}(t',t) \Big].
\end{multline}
Averaging over stochastic process makes current time-independent  and moving to Fourier space gives
\begin{equation}
\langle J_L \rangle_\xi = 
2\;    \text{Re} \int  \frac{d \omega}{2 \pi}  \text{Tr} \Big[  \bm {\mathcal G}^<(\omega) \bm \Sigma^A_{\alpha}(\omega)
+  \bm {\mathcal G}^R(\omega)  \bm \Sigma^<_{\alpha}(\omega) \Big].
\end{equation}
Making assumption that  imaginary parts of retarded or advanced  self-energies  for left and right electrodes are proportional to each other,
the average current can be expressed only in terms of  retarded Green's function as
\begin{equation}
 \langle J_L \rangle_\xi
=     \int \frac{d \omega}{2 \pi} T(\omega) (f_L(\omega) -f_R(\omega)),
\end{equation}
where
\begin{equation}
T(\omega)= -\frac{1}{ \pi} \text{Tr} \Big[   
\frac{\bm \Gamma_L(\omega) \bm \Gamma_R(\omega)}{\bm \Gamma_L(\omega) + \bm \Gamma_R(\omega)}  
\text{Im} \{ \bm {\mathcal G}^R(\omega) \} \big]
\label{T1}
\end{equation}
is the transmission coefficient averaged over the stochastic process and  $\bm {\mathcal G}^R(\omega) $ is given by equation (\ref{gr}).

\section{Applications}

\subsection{Fluctuating resonant-level}

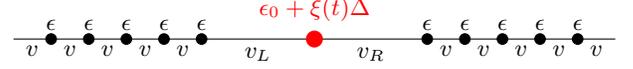
\begin{figure}
\begin{tikzpicture}
\draw (-4,0) -- (4,0);

\filldraw [black] (-1.5,0) circle (2pt) node[above] {$\epsilon$};
\node[] at (-1.75,-0.15)  {$v$};
\filldraw [black] (-2.0,0) circle (2pt) node[above] {$\epsilon$};
\node[] at (-2.25,-0.15)  {$v$};
\filldraw [black] (-2.5,0) circle (2pt) node[above] {$\epsilon$};
\node[] at (-2.75,-0.15)  {$v$};
\filldraw [black] (-3.0,0) circle (2pt) node[above] {$\epsilon$};
\node[] at (-3.25,-0.15)  {$v$};
\filldraw [black] (-3.5,0) circle (2pt) node[above] {$\epsilon$};
\node[] at (-3.75,-0.15)  {$v$};

\node[] at (-0.75,-0.2)  {$v_L$};
\filldraw [red] (0,0) circle (3pt) node[] at (0, 0.4)  {$\epsilon_0 + \xi(t) \Delta$};
\node[] at (0.75,-0.2)  {$v_R$};

\filldraw [black] (1.5,0) circle (2pt) node[above] {$\epsilon$};
\node[] at (1.75,-0.15)  {$v$};
\filldraw [black] (2.0,0) circle (2pt) node[above] {$\epsilon$};
\node[] at (2.25,-0.15)  {$v$};
\filldraw [black] (2.5,0) circle (2pt) node[above] {$\epsilon$};
\node[] at (2.75,-0.15)  {$v$};
\filldraw [black] (3.0,0) circle (2pt) node[above] {$\epsilon$};
\node[] at (3.25,-0.15)  {$v$};
\filldraw [black] (3.5,0) circle (2pt) node[above] {$\epsilon$};
\node[] at (3.75,-0.15)  {$v$};

\end{tikzpicture}
\caption{Single fluctuating resonant level connected to  left and right electrodes represented by two semi-infinite one-dimensional monoatomic chains. }
\label{fig1}
\end{figure}

\begin{figure}
\centering
\includegraphics[width=1\columnwidth]{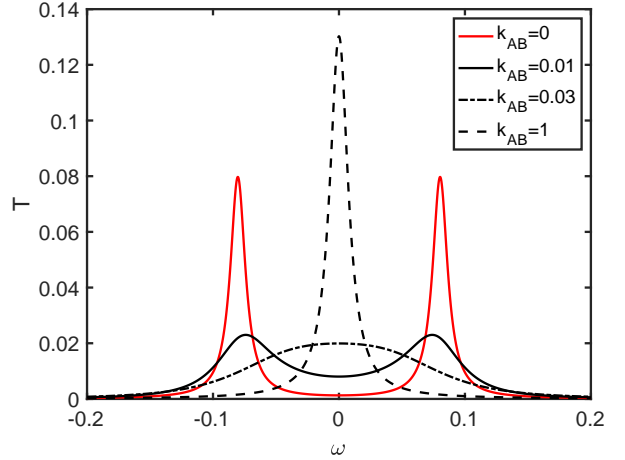}
\caption{Transmission coefficient computed for different rates $k_{AB}$ of stochastic switching, values of $k_{AB}$ are given in units of $\Gamma$. Parameters used in calculations (in a.u.): $\epsilon_0=0$, $\Delta =0.04$, $v_L=v_R=0.03$, $v=0.5$.}
\label{fig2}
\end{figure}

Consider electron transport through a single fluctuating resonant level shown in Fig.\ref{fig1}. We take molecular Hamiltonian in the following form
\begin{equation}
H_M(t) = (\epsilon + \xi(t) \Delta  ) a^\dag a,
\end{equation}
and the  self-energy  is computed using one-dimensional monoatomic chain as electrode model (appendix A).
The retarded Green's function (\ref{gr}) becomes
\begin{equation}
 {\mathcal G}^{R}(\omega)=\Big[  \omega -  \epsilon  -\Lambda + \frac{i}{2} \Gamma  -  \frac{\Delta^2}{\omega -\epsilon -\Lambda + \frac{i}{2} \Gamma+ 2 i k_{AB}    }   \Big]^{-1}
 \label{gr1l}
\end{equation}

Fig.\ref{fig2} shows transmission coefficient computed using eq.(\ref{T1}) with retarded Green's function ({\ref{gr1l}). If switching rate $k_{AB}$ is set to zero, then the transmission coefficient has two resonance peaks corresponding to energies  $\epsilon + \Delta \epsilon $ and $\epsilon - \Delta \epsilon $. Once the level starts to fluctuate between two values, the broadening of individual stochastic realization of resonance level increases at first, this behavior is expected. It resembles the electron transport through a single level with delta-correlated noise where delta-correlated fluctuations of resonance level energy results in an additive contribution to $\Gamma$ \cite{rubo93}. However, when the switching rate becomes comparable with the rate of electron transport,  two resonances merge into a single energetically broad transport channel. The further increase of the switching rate narrows this initially broad transport channel in a narrow single transmission resonance.

\subsection{Stochastic forming and breaking of  bond }

We consider molecular electronic junctions with two fluctuating bonds which separate one part of the molecule from another. One may view this as an example of a molecular junction sensor to analyze the bonding dynamics with some environmental molecules. Fig.\ref{fig3} sketches the model. 

The static part of the  molecular Hamiltonian matrix is
\begin{equation}
\bm H =
\begin{bmatrix}
\epsilon_0 & \beta         & 0               & 0                & 0              & 0                 &0               \\
0               & \epsilon_0 & \beta         & 0                & 0              & 0                &\beta         \\
0               & \beta         & \epsilon_0 & \frac{1}{2} \beta          & 0              & 0                &0              \\
0               & 0               & \frac{1}{2} \beta          & \epsilon_0 & \beta         & 0                &0              \\
0               &0                & 0               & \beta          & \epsilon_0 &\beta           &0             \\
0               & 0               & 0               & 0                & \beta          & \epsilon_0 &\frac{1}{2}\beta               \\
0               & \beta         & 0               & 0                & 0                &\frac{1}{2} \beta         &\epsilon_0 \\
\end{bmatrix},
\end{equation}
and the fluctuating part is given by the following expression
\begin{equation}
\bm V =
\begin{bmatrix}
0               &0                & 0               & 0                & 0              & 0                 &0               \\
0               &0                & 0                & 0                & 0              & 0                &0        \\
0               &0                 & 0               &\frac{1}{2} \beta          & 0              & 0                &0              \\
0               & 0               &\frac{1}{2}  \beta          &0                & 0             & 0                &0              \\
0               &0                & 0               & 0              & 0                &0           &0             \\
0               & 0               & 0               & 0                &0                  & 0              &\frac{1}{2} \beta               \\
0               &0                & 0               & 0                & 0                & \frac{1}{2} \beta         &0 \\
\end{bmatrix}.
\end{equation}
Here $\epsilon_0$ is the site energy and $\beta$ is the hopping amplitute. The numbering of atoms shown in Fig.\ref{fig3} corresponds to ordering of matrix elements in the above matrices. The hopping amplitude for fluctuating bond in Fig.\ref{fig3}
switches between 0  and $\beta$, which corresponds to bond dissociation and formation, respectively.

\begin{figure}
\begin{tikzpicture}

\draw (-4,0) -- (4,0);
\filldraw [black] (-1.5,0) circle (2pt) node[above] {$\epsilon$};
\node[] at (-1.75,-0.15)  {$v$};
\filldraw [black] (-2.0,0) circle (2pt) node[above] {$\epsilon$};
\node[] at (-2.25,-0.15)  {$v$};
\filldraw [black] (-2.5,0) circle (2pt) node[above] {$\epsilon$};
\node[] at (-2.75,-0.15)  {$v$};
\filldraw [black] (-3.0,0) circle (2pt) node[above] {$\epsilon$};
\node[] at (-3.25,-0.15)  {$v$};
\filldraw [black] (-3.5,0) circle (2pt) node[above] {$\epsilon$};
\node[] at (-3.75,-0.15)  {$v$};

\node[] at (-0.75,-0.2)  {$v_L$};
\filldraw [red] (0,0) circle (2pt) node[below] {1};
\node[] at (0.75,-0.2)  {$v_R$};
\draw[red] (0,0) -- (0,0.5);
\filldraw [red] (0,0.5) circle (2pt) node[above]{2};

\draw[red]  (0,0.5) -- ({0.5*cos(30)}, {0.5+0.5*sin(30)});
\filldraw [red] ({0.5*cos(30)}, {0.5+0.5*sin(30)}) circle (2pt) node[below] {7};
\draw[red]  (0,0.5) -- ({-0.5*cos(30)}, {0.5+0.5*sin(30)});
\filldraw [red] ({-0.5*cos(30)}, {0.5+0.5*sin(30)}) circle (2pt) node[below] {3};

\draw[red,dotted, line width = 1.5pt]   ({0.5*cos(30)}, {0.5+0.5*sin(30)}) -- ({0.5*cos(30)}, {0.5+0.5+0.5*sin(30)});
\filldraw [red] ({0.5*cos(30)}, {0.5+0.5+0.5*sin(30)}) circle (2pt) node[above]{6};
\draw[red,dotted, line width = 1.5pt]   ({-0.5*cos(30)}, {0.5+0.5*sin(30)}) -- ({-0.5*cos(30)}, {0.5+0.5+0.5*sin(30)});
\filldraw [red] ({-0.5*cos(30)}, {0.5+0.5+0.5*sin(30)}) circle (2pt) node[above]{4};

\draw[red]   ({0.5*cos(30)}, {0.5+0.5+0.5*sin(30)}) -- (0.0, {0.5+0.5+0.5*sin(30)+0.5*sin(30)});
\draw[red]   ({-0.5*cos(30)}, {0.5+0.5+0.5*sin(30)}) -- (0.0, {0.5+0.5+0.5*sin(30)+0.5*sin(30)});
\filldraw [red] (0.0, {0.5+0.5+0.5*sin(30)+0.5*sin(30)}) circle (2pt) node[above]{5};

\filldraw [black] (1.5,0) circle (2pt) node[above] {$\epsilon$};
\node[] at (1.75,-0.15)  {$v$};
\filldraw [black] (2.0,0) circle (2pt) node[above] {$\epsilon$};
\node[] at (2.25,-0.15)  {$v$};
\filldraw [black] (2.5,0) circle (2pt) node[above] {$\epsilon$};
\node[] at (2.75,-0.15)  {$v$};
\filldraw [black] (3.0,0) circle (2pt) node[above] {$\epsilon$};
\node[] at (3.25,-0.15)  {$v$};
\filldraw [black] (3.5,0) circle (2pt) node[above] {$\epsilon$};
\node[] at (3.75,-0.15)  {$v$};

\end{tikzpicture}
\caption{Molecular junction with on and off fluctuating bonds (dashed line).  The left and right electrodes are represented by two semi-infinite one-dimensional monoatomic chains.
  }
\label{fig3}
\end{figure}
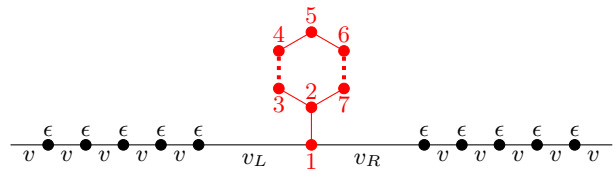

\begin{figure}
\centering
\includegraphics[width=1\columnwidth]{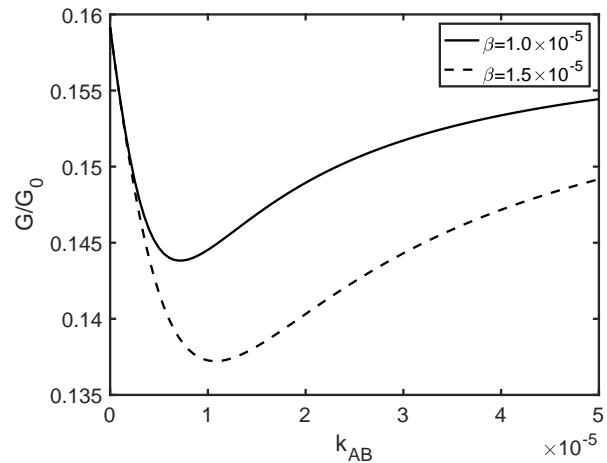}
\caption{Conductance  as a function of switching rate. $G_0$ is the conductance quantum.  Parameters used in calculations: $\epsilon_0=0$,  $v_L=v_R=0.001$, $v=0.5$.}
\label{fig4}
\end{figure}

Fig.4 shows the conductance of the molecular junction as a function of the switching rate computed for different intramolecular hopping amplitudes
If  $k_{AB}=0$, electron travels through the  static junction. The resulting conductance is the average of conductances for geometry A and geometry B,
$\frac{1}{2}(G_A + G_B)$, due to our  assumption that the initial configuration has an equal chance of being A or B.  
Once the switching rate becomes non-zero, the conductance value drops first, reaching the minimum when $k_{AB}$ is the same order of magnitude as intermolecular hopping $\beta$. The further switching rate increases lead to the conductance increase with subsequent plateau.

\section{Conclusions}

We have developed a theory that includes dynamics of molecular stochastic structural changes in nonequilibrium Green's functions electron transport calculations. 
In our theory, we assumed that the molecular hamiltonian switches stochastically at a random time between two different operators. Furthermore, we took the distribution of times at which the transitions occur to be the Poisson stationary distribution with a constant average number of transitions per unit of time. These are standard assumptions of stochastic telegraph processes.

We derived equations of motion for stochastically averaged retarded Green's function in closed form. The closure requires the additional differential equation for the auxiliary quantity $\langle \xi(t) \bm G(t, t') \rangle_\xi$, which we obtained using the Novikov-Furutsu method of functional stochastic calculus. Next, we averaged electric current over time and received an expression for the transmission coefficient, which depends on the rates for structural molecular transitions. 

The proposed method has a  limitation - we had to assume that the imaginary parts of the left and right electrodes' retarded self-energies are proportional to each other. This limitation significantly restricts possible molecular junction geometries to which we can apply the theory.

We illustrated the theory first by considering a simple model of electron transport through a single fluctuating current carrying molecular orbitale. Next,  we studied the electrical properties of the supramolecular model complex with fluctuating links separating two structural parts.

\clearpage 
\begin{widetext}
\appendix
\section{Self-energy}
The real and imaginary parts computed  for monoatomic semi-infinite chain with site energy $\epsilon$ and hopping interaction $v$ are
\begin{eqnarray}
\Lambda_\alpha(\omega) =
\frac{|v_\alpha|^2}{2 |v|^2} \times
\left\{ 
\begin{array}{cc}
  \omega -\epsilon - \sqrt{(\omega- \epsilon)^2 - 4 v^2}, & \text{ if }  \omega -\epsilon \ge 2 |v|
\\
\\
 \omega -\epsilon , & \text{ if }  |\omega -\epsilon| <  2 |v|
\\
\\
\omega -\epsilon + \sqrt{(\omega- \epsilon)^2 - 4 v^2}, & \text{ if }  \omega -\epsilon \le -2 |v|
\end{array}
\right.
\end{eqnarray}
\begin{eqnarray}
\Gamma_\alpha(\omega) = \frac{|v_\alpha|^2}{|v|^2}  \times
\left\{ 
\begin{array}{cc}
0,  \text{ if }  \omega -\epsilon \ge 2 |v|
\\
\\
 \sqrt{4 v^2 - (\omega- \epsilon)^2 }, & \text{ if }  |\omega -\epsilon| <  2 |v|
\\
\\
0, & \text{ if }  \omega -\epsilon \le -2 |v|
\end{array}
\right.,
\end{eqnarray}

\section{Novikov-Furutsu functional approach to stochastic averaging}

Following  ideas of  Novikov-Furutsu theory \cite{novikov64,furutsu63} we introduce
\begin{equation}
\Big \langle \xi(t)    \bm    G^R[\xi+z](t,t') \Big\rangle_\xi,
\end{equation}
where $z$ is an arbitrary reasonably smooth deterministic function.
We expand $ \bm    G^A[\xi+z](t,t')$ in the Taylor-like series using functional differentiation around $\xi(t)=0$
\begin{equation}
  \bm    G^R[\xi+z](t,t') =   \bm    G^R[z](t,t') + \int_{-\infty}^t dt_1 \frac{\delta   \bm    G^R[z](t,t')}{\delta z(t_1)} \xi(t_1) 
+ \frac{1}{2!}\int_{-\infty}^{t} dt_1 \int_{-\infty}^{t} dt_2 \frac{\delta^2   \bm    G^R[z](t,t')}{\delta z(t_1) \delta z(t_2)} \xi(t_1) \xi(t_2)  + ....,
\end{equation}
which can be compactly written  with the use of the operator exponent
as
\begin{equation}
 \bm    G^R[\xi+z](t,t') = \exp \Big\{ \int_{-\infty}^t dt_1 \xi(t_1) \frac{\delta}{\delta z(t_1)} \Big\} \;   \bm    G^R[z](t,t').
\end{equation}
Notice that the above derivations have subtly assume that the retarded Greens $G^R[z](t,t')$ is a function of stochastic process $\xi(t_1)$ for $t_1\le t$ and it does not depend on  $\xi(t_1)$ for $t_1 > t$. It follows from the presence of step function $\theta(t-t')$ in the definition of the retarded Green's function (\ref{gr}).
The similar calculations can be performed for $G^A[z](t,t')$ using $t'$ as a reference time for the functional shift. However, it is not clear to us  how to perform 
similar derivations for lesser Green's function.

Let us now compute
\begin{equation}
 \Big\langle \xi(t)    \bm    G^R[\xi+z](t,t')  \Big\rangle_\xi
 =  \Big\langle \xi(t)  \exp \Big\{ \int_{-\infty}^t dt_1 \xi(t_1) \frac{\delta}{\delta z(t_1)} \Big\}  \Big\rangle_\xi \;    \bm    G^R[z](t,t').
\end{equation}
Expanding the operator exponent, we get
\begin{multline}
\Big\langle \xi(t)    \bm    G^R[\xi+z](t,t') \Big\rangle_\xi= \\
  \Big\langle \xi(t)  \Big[1+ \sum_{k=1}^\infty \frac{1}{k!} \int_{-\infty}^t dt_1 \int_{-\infty}^t dt_2 .... \int_{-\infty}^t dt_k \xi(t_1) \xi(t_2) ... \xi(t_k) 
\frac{\delta}{\delta z(t_1)} \frac{\delta}{\delta z(t_2)} ... \frac{\delta}{\delta z(t_k)} \Big] \Big\rangle_\xi \;    \bm    G^R[z](t,t') =
\\
  \Big[\Big\langle \xi(t) \Big\rangle_\xi +  \sum_{k=1}^\infty  \int_{-\infty}^t dt_1 \int_{-\infty}^{t_1} dt_2 .... \int_{-\infty}^{t_{k-1}} dt_k    \Big\langle \xi(t) \xi(t_1) \Big\rangle_\xi   \Big\langle  \xi(t_2) ... \xi(t_k)  \Big\rangle_\xi 
\frac{\delta}{\delta z(t_1)}\frac{\delta}{\delta z(t_2)} ... \frac{\delta}{\delta z(t_k)} \Big]  \;    \bm    G^R[z](t,t') 
\\
=   \Big[ \Big\langle \xi(t) \Big\rangle_\xi +    \int_{-\infty}^t dt_1    \Big\langle \xi(t) \xi(t_1) \Big\rangle_\xi \frac{\delta}{\delta z(t_1)}
\Big\langle \exp \Big\{ \int_{-\infty}^{t_1} dt' \xi(t') \frac{\delta}{\delta z(t')} \Big\} \Big\rangle_\xi \Big]  \;    \bm    G^R[z](t,t').
\end{multline}
Substituting expression for moments of stochastic process yields
\begin{equation}
\label{a6}
\langle \xi(t)    \bm    G^R[\xi+z](t,t') \rangle_\xi
= \Big[       \int_{-\infty}^{t} dt_1    e^{-2 k_{AB} (t-t_1)} \frac{\delta}{\delta z(t_1)}
\Big\langle \exp \Big\{ \int_{-\infty}^{t_1} dt_2 \xi(t_2) \frac{\delta}{\delta z(t_2)} \Big\} \Big\rangle_\xi \Big]  \;    \bm    G^R[z](t,t'),
\end{equation}

Differentiating the lhs and rhs of this equation with respect to $t$ gives
\begin{equation}
(\partial_t + 2 k_{AB}) \langle \xi(t)    \bm    G^R[\xi+z](t,t') \rangle_\xi = \Big[       \int_{-\infty}^{t} dt_1    e^{-2 k_{AB} (t-t_1)} \frac{\delta}{\delta z(t_1)}
\Big\langle \exp \Big\{ \int_{-\infty}^{t_1} dt_2 \xi(t_2) \frac{\delta}{\delta z(t_2)} \Big\} \Big\rangle_\xi \Big]  \;   \partial_t  \bm    G^R[z](t,t')
\end{equation}
Notice that when computing derivative we should not differentiate the time limit in the integral\cite{klyatskin2011}.
Comparing rhs of the equation with (\ref{a6}) we get
\begin{equation}
(\partial_t + 2 k_{AB}) \langle \xi(t)    \bm    G^R[\xi+z](t,t') \rangle_\xi =    \langle \xi(t)  \;\partial_t   \bm    G^R[\xi+z](t,t') \rangle_\xi.
\end{equation}
Setting $z(t)=0$ in the above equation gives
\begin{equation}
(\partial_t + 2 k_{AB}) \langle \xi(t)    \bm    G^R[\xi](t,t') \rangle_\xi =    \langle \xi(t)  \;\partial_t   \bm    G^R[\xi](t,t') \rangle_\xi.
\end{equation}
This equation for time derivative of product of a stochastic functional  with corresponding stochastic process is  sometimes called the  Shapiro-Loginov differential formula \cite{shapiro-loginov}.

\end{widetext}

\clearpage
\nocite{}
\bibliography{../../../../Documents/BIBLIOGRAPHY/thphys.bib}

\end{document}